# Thermally-driven Multilevel Non-volatile Memory with Monolayer MoS$_2$ for Neuro-inspired Artificial Learning

Sameer Kumar Mallik,[1,2] Roshan Padhan,[1,2] Mousam Charan Sahu,[1,2] Suman Roy,[1,2] Gopal K Pradhan,[3] Prasana Kumar Sahoo,[4] Saroj Prasad Dash,[5] Satyaprakash Sahoo[1,2]*

[1]*Laboratory for Low Dimensional Materials, Institute of Physics, Bhubaneswar-751005, India*
[2]*Homi Bhabha National Institute, Training School Complex, Anushakti Nagar, Mumbai, 400094, India*
[3]*Department of Physics, School of Applied Sciences, KIIT Deemed to be University, Bhubaneswar, Odisha 751024, India*
[4]*Materials Science Centre, Quantum Materials and Device Research Laboratory, Indian Institute of Technology Kharagpur, Kharagpur, West Bengal, India*
[5]*Quantum Device Physics Laboratory, Department of Microtechnology and Nanoscience, Chalmers University of Technology, Göteborg, Sweden*

## Abstract

The demands of modern electronic components require advanced computing platforms for efficient information processing to realize in-memory operations with a high density of data storage capabilities towards developing alternatives to von Neumann architectures. Herein, we demonstrate the multifunctionality of monolayer MoS$_2$ mem-transistors which can be used as a high-geared intrinsic transistor at room temperature; however, at a high temperature (>350 K), they exhibit synaptic multi-level memory operations. The temperature-dependent memory mechanism is governed by interfacial physics, which solely depends on the gate field modulated ion dynamics and charge transfer at the MoS$_2$/dielectric interface. We have proposed a non-volatile memory application using a single FET device where thermal energy can be ventured to aid the memory functions with multi-level (3-bit) storage capabilities. Furthermore, our devices exhibit linear and symmetry in conductance weight updates when subjected to electrical potentiation and depression. This feature has enabled us to attain a high classification accuracy while training and testing the Modified National Institute of Standards and Technology datasets through artificial neural network simulation. This work paves the way for new avenues in 2D semiconductors toward reliable data processing and storage with high-packing density arrays for brain-inspired artificial learning.

*Corresponding Author: sahoo@iopb.res.in

# 1. Introduction

In this modern era of electronic devices, ultrahigh data storage density is imperative for smarter computing platforms and vigorous information processing applications. Owing to the physical limits of silicon-based metal oxide field effect transistors (MOSFETs), the quest for alternative prospective materials has been the focus of intensive research to improve storage capacity. The discovery of atomically thin graphene has put forward direction toward significant active research in other two-dimensional (2D) materials over the last decade.[1] Molybdenum disulfide ($MoS_2$),[2] being an in-demand member of the transition metal di-chalcogenides (TMDCs) family,[3] has met broadband applications in electronics and optoelectronics manifesting its potency for future miniaturized electronic components.[4,5] Atomically thin $MoS_2$ channel based FETs exhibit high carrier mobility ~200 $cm^2V^{-1}s^{-1}$, and profound ON/OFF switching characteristics ~$10^8$.[6] However, its large surface-to-volume ratio imposes an inevitable exposure of the atomically thin channel to the oxide trap/defect states, which draw hysteresis and threshold voltage instabilities hindering device performances.[7,8] Nevertheless, the large hysteresis produced during the dual sweep and pulsed gate operations has recently been exploited for nonvolatile (which include 2D flash memory,[9] magnetic random access memory,[10] resistive random access memory[11,12]) and volatile (such as dynamic random access memory,[13] semi-floating gate transistor[14]) memory applications. Considering the superior charge trap capacity of conventional $Al_2O_3/HfO_2/Al_2O_3$ gate stack and few layer $MoS_2$ as the channel, Zhang *et. al.* establish a large memory window (~20 V) with efficient modulation of the program and erase operation.[15] A stable two-terminal floating gate memory cell is realized by Vu *et. al.* on a vertically stacked $MoS_2$/hBN/graphene heterostructure providing program/erase operations with retention >$10^4$ s.[16] However, cumbersome and exorbitant processes are still demanded for making charge-trap based memory devices. Additionally, the size of the overall devices increases with multiple gate stacking and various tailor-made designs. To achieve higher storage density per unit cell area, growing attention has spurred the creation of ultra-thin 2D materials based multibit memory devices.

Recently, memtransistors (integration of transistors with memristive functionalities) based on 2D MoS$_2$ have been proposed with various architectures showing controlled transport with digital switching ratios,[17,18] multibit optoelectronic memory,[19] neuromorphic computing applications etc.[20,21] Especially, the electrolyte gated transistors show better conductance modulation benefitted from its ion-gating mechanism over electrostatic charge trap phenomena, making them viable candidates for brain-inspired computation and logic-in-memory applications.[22,23] Despite the recent development of synaptic ion-gated transistors, multi-level memory based on ion-gating mechanisms is still lacking. However, liquid electrolytes practically limit the large-density integration of devices and high-temperature applications. Consequently, it is desirable to instigate new multi-level memories with eccentric functionalities that confer the possibility of robust device architecture with higher throughput. Furthermore, with the growing packing density of FET arrays on a single wafer, high performance integrated circuits (ICs) can reach an operating temperature as high as 500K,[24] making it essential to understand and exploit novel properties of 2D materials based devices at high temperatures. Thermally-driven memory is one of the applications where thermal energy can aid memory functions with multi-level storage capabilities.

Our work demonstrates robust MoS$_2$/SiO$_2$/Si three-terminal devices with very different functionality and a novel mechanism that drives to a high geared intrinsic transistor at room temperature and exhibits a multi-bit memory cell above room temperature. The proposed thermally-driven programmable memory operations, i.e., READ, WRITE, ERASE, are found to be modulated by gate voltage biasing history. The step-like READ-RESET ratio and retention curves are obtained for several pulse voltage conditions. The multi-level states are investigated with varying pulsed width, gate voltage amplitudes, and drain/source voltages. We proposed the energy band diagrams that explain the single and multi-level memory effects in MoS$_2$ FETs to validate the results. Furthermore, we demonstrate excellent linearity and symmetry upon electrical potentiation and depression. As a result, a high classification accuracy is achieved during training and testing of the Modified National Institute of Standards and Technology (MNIST) datasets in artificial neural network (ANN) simulation. These findings indicate the potential of in-memory applications in ion-driven MoS$_2$-based transistors for building multilevel memory and synapse arrays, facilitating complex data processing tasks.

## 2. Results and Discussion

Monolayer (ML) MoS$_2$ (triangular domains of size ~30 $\mu$m) are directly synthesized on SiO$_2$(285 nm)/Si substrate by using the salt-assisted chemical vapor deposition (CVD) technique (see the Methods section for more details).[25] Figure 1(a) displays the prominent Raman active modes i.e. E$_{2g}$ and A$_{1g}$ of as-grown MoS$_2$. The frequency difference ($\Delta\omega$= 19 cm$^{-1}$) between these phonon modes are characteristic signature of ML nature of our sample. The Lorentz fit results provide the FWHM of E$_{2g}$ (~2.5 cm$^{-1}$) and A$_{1g}$ (~5.2 cm$^{-1}$) modes comparable to that of exfoliated MoS$_2$ which depict high crystallinity of our CVD grown samples.[26] Figure 1(b) displays the optical micrograph of as-fabricated MoS$_2$ devices with silver (Ag) as drain/source electrodes. The near equal work function of Ag and electron affinity of ML MoS$_2$ allows the formation of a low Schottky barrier with negligible strain effect and superior carrier transport when Ag is used as metal contact to ML MoS$_2$.[25,27] The channel lengths (L$_{ch}$) of proposed devices are varied from 1 to 4 $\mu$m with channel width fixed at 10 $\mu$m. Photoluminescence (PL) mapping of the MoS$_2$ device shown in Fig. 1(c) provides intense and uniform luminescence, which illustrates exemplary optical properties of the ML MoS$_2$ throughout the channel regions.

Initially, the room temperature basic transistor properties are obtained for L$_{ch}$ = 2 $\mu$m in a high vacuum environment (10$^{-6}$ mbar) under dark conditions. Figure 1(d) demonstrates the linear output characteristics (i.e., drain current $I_d$ vs. drain-source voltage $V_{ds}$) for the $V_{ds}$ sweep range from -100 to 100 mV at fixed back gate voltages ($V_g$ = 0 to 40V in steps of 5V). Linear $I_d$ vs. $V_{ds}$ curves and $I_d$ values in few $\mu$A ensure the Ohmic contact has been formed between Ag and $MoS_2$,[28] which drives excellent carrier transport with low power consumption as per the previous literature.[25] The stable electrostatic gate control with n-type transistor behavior can be observed in Fig. 1 (e) during the single sweep transfer characteristics ($I_d$ vs. $V_{ds}$) for $V_g$ sweep range from -40 to 40V with fixed $V_{ds}$ values (25, 50, 75, and 100 mV). The semi-logarithmic plot shows a high on/off current ratio ($I_{on}/I_{off}$) of about $10^5$ providing excellent transistor switching operations. The field effect mobility ($\mu_{FE}$) and threshold voltage ($V_{th}$) is extracted from the slope of the linear region of the transfer curve for $V_{ds}$ = 100 mV,[25] which is shown in Fig. 1(f). In our case, such high carrier mobility of ~22 cm$^2$V$^{-1}$s$^{-1}$ is free from grain boundary and strain effects and therefore meets excellent agreement with previously reported results on unpassivated ML $MoS_2$.[25,29] The transistor properties of other channel lengths are provided in the Supplementary Information (see Figure S1). The as-prepared backgated $MoS_2$ devices with Ag as metal electrodes show superior transistor performance at room temperature and are further investigated for temperature-dependent behavior.

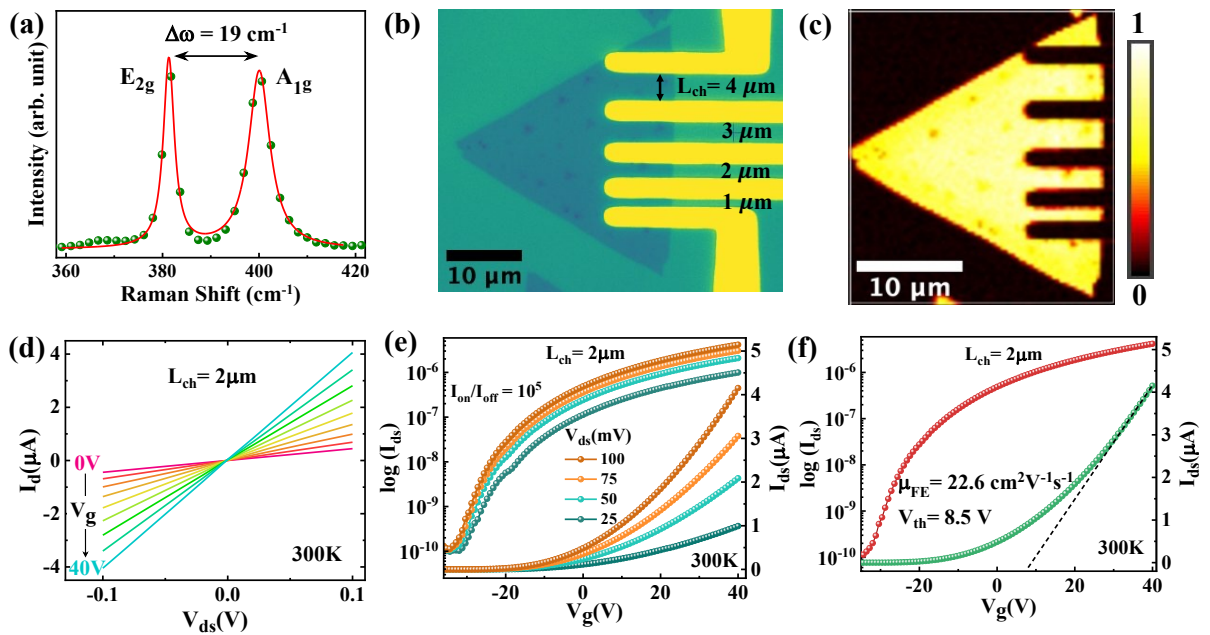

***Figure 1. Structural characterization, design, and electrical properties of MoS$_2$ transistor***

*(a) The Raman spectroscopic signature of ML MoS$_2$ show E$_{2g}$ and A$_{1g}$ modes. The Lorentz fit results (solid red lines) provide the high crystallinity of CVD grown ML MoS$_2$ sample. (b) Optical image of the MoS$_2$ device having equal channel widths (10 μm) and length varies from 1 μm to 4 μm, Ag is being used as source/drain electrodes (shown in false color). (c) Photoluminescence mapping confirms the uniform optical grade quality of the MoS$_2$ sample with robust device fabrication. (d) The output characteristics (I$_d$ vs V$_{ds}$) for fixed gate voltages (V$_g$ ranging from 0V to 40 V) show the Ohmic contact nature of the as fabricated device. (e) The transfer characteristics (I$_d$ vs V$_g$) is displayed in both linear and semi-logarithmic scale for fixed drain-source voltages (25 mV to 100 mV) showing stable electrostatic gate control and the current on/off ratio ~ 10$^5$. (f) The field-effect mobility (μ$_{FE}$) and threshold voltage (V$_{th}$) is extracted from the slope of the linear region of the transfer curve.*

The temperature-dependent dual sweep transfer characteristics are conducted to perceive the evolution of hysteresis in as-fabricated devices under high vacuum (~10$^{-6}$ mbar). The hysteresis behavior over the temperature range from 100 to 450 K with gate voltage forward sweep from -60 to 40 V and backward sweep from 40 to -60 V for different channel lengths are plotted in Fig. S2. The indiscernible dual sweep curves from 100 to 400 K stipulate weak hysteretic behavior manifesting lessened gate stress effects, unvaried threshold voltages during gate operations,[8] and hence elicit excellent transistor performance. Nevertheless, noticeable hysteresis is observed above 400 K, which inflates further at elevated temperatures. According to previous reports, the origin of hysteresis in MoS$_2$ based FETs can have various source factors at room temperature, such as adsorption of oxygen/water molecules onto MoS$_2$ channel surface at high bias in ambient pressure,[7] charge trapping/detrapping at MoS$_2$-dielectric interface, etc.[30] However, at high temperatures, the charge trap density increases due to the thermal activation of deep trap states in the dielectric, allowing a faster trapping process which essentially causes outspread hysteresis.[30] A recent report by He *et. al.* explains that a charge injection from the back gate to dielectric that drives unique hysteretic crossover characteristics at relatively higher temperatures.[31] Kaushik *et. al.* describe independent mechanisms occurring at room and high temperatures in the case of FETs based on multilayered exfoliated $MoS_2$, that switches the hysteresis loop from clockwise to anticlockwise.[32]

The room temperature transfer characteristics in our case show robust transistor performance due to its minimal hysteresis behavior. A transfer curve at temperature 250K and 300K for $L_{ch}$ = 2 $\mu$m is shown in Fig. 2 (a), along with possible mechanisms illustrated in Fig. 2 (b) that justify the nature of the hysteresis loop. Low backward sweep current (BSC) indicates clockwise hysteresis in a dual sweep transfer curve, whereas high BSC reflects anti-clockwise hysteretic nature. The red (blue) curve signifies a forward (backward) sweep to realize the clockwise hysteresis nature at low and room temperature having hysteresis width less than 1V. As shown in Fig. 2 (b), two cross-sectional schematic representations of the device portray the possible interfacial mechanism that involves charge-trapping processes at high negative and positive gate voltages. As illustrated in the schematic, at the beginning of the forward sweep, when a particular negative bias is applied to the gate terminal, an initial de-trapping process in the $SiO_2$ releases additional electrons to the $MoS_2$ channel, which causes a left shift of the transfer characteristics. As we reach a high positive gate bias, the capturing of electrons at the trapping sites of $SiO_2$ dominates. As a consequence, during the backward sweep, the channel of $MoS_2$ contains fewer electrons, which leads to low BSC.[25]

The lower panel of Fig. 2 (a) shows the transfer characteristics at a high temperature (~ 450 K), depicting a high BSC and hence anti-clockwise hysteresis behavior with a large voltage window. The earlier results on the anti-clockwise hysteresis in the case of $MoS_2$ FETs are reported at high temperature where a completely different form of hysteresis has been observed, associated with a sudden current jump near zero gate bias.[31,32] He *et. al.* proposed a distinct mechanism that entails the charge injection from the gate (n-type Si) to the thermally activated deep trap states in $SiO_2$, creating a repulsive and attractive gate field depending on the bias polarization.[31] However, the nature of the anti-clockwise hysteresis in our case is distinctly different from that of previous reports and requires an independent mechanism that can corroborate our forthcoming results. The fundamental idea that propels the origin of anti-clockwise hysteresis in our case is as follows; in NaCl-assisted CVD growth of $MoS_2$, the minute amount of mobile charges such as sodium ions ($Na^+$) exist in the bulk of $SiO_2$ gate dielectric is inevitable during high temperature synthesis process. It is apparent that the mobile ions in gate oxide can be a plausible factor responsible for threshold voltage shift and could induce hysteretic transfer characteristics in FETs.[33] As shown in schematic of Fig. 2(c), under the application of high negative bias at the gate terminal, the $Na^+$ ions move away from the channel-interface region, and a random distribution of ions can be realized in the gate dielectric during the forward sweep. However, under high positive bias, these ions drift towards the $MoS_2$ channel, which eventually increases the drain current values during the backward sweep. We reproduce the above results on evolution of hysteresis by fabricating devices with different channel lengths ($L_{ch}$ = 1, 3, and, 4 μm) and the obtain results are shown in Fig. S3. Even with different channel lengths the repeatability of our results are highly reproducible over the temperature range from 100 to 450 K indicating the high temperature anti-clockwise hysteresis behavior is not limited to a confined channel length. To support our $Na^+$ ion-driven mechanism, it is extremely important to compare these results with conventional (without NaCl) CVD grown $MoS_2$ based transistors. For this, we have conducted the dual sweep transfer characteristics on such CVD grown pristine $MoS_2$ samples to examine the role of $Na^+$ ions on device performances. It is worth mentioning that the hysteresis behavior of without NaCl CVD grown $MoS_2$ based transistors show a consistent clockwise behavior throughout the temperature range from 250 to 450 K, as shown in Fig. 2 (d) and supplementary Fig. S4. This behavior is attributed to the increased thermally activated trap centers at MoS$_2$/SiO$_2$ interface which trap/de-trap charge carriers depending on the voltage polarization. This comparison confirms our proposed mechanism of the Na$^+$ ion-driven high temperature anti-clockwise hysteresis in MoS$_2$ transistors. In the subsequent sections, we shall expound the mechanism more lucidly and observe how the ions migrate inside SiO$_2$, generating a diffusion-like drain current behavior in the conduction channel with temperature.

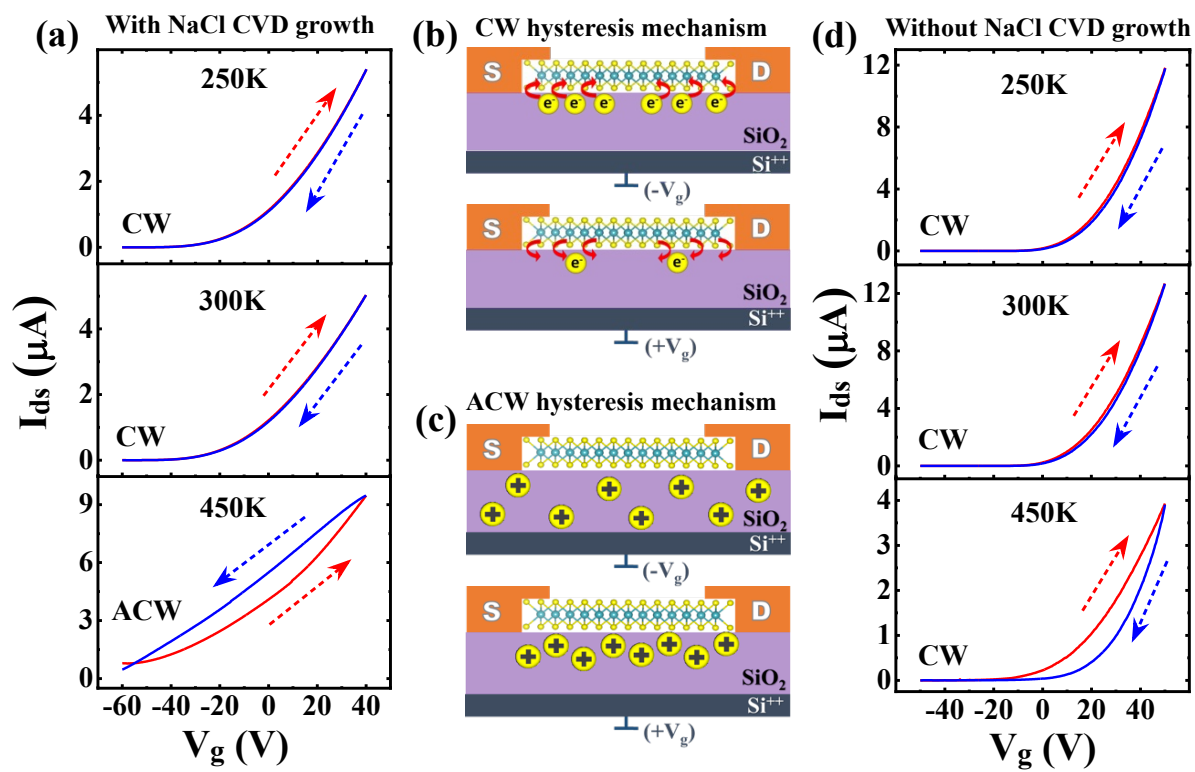

*Figure 2. Temperature-dependent hysteresis evolution and proposed mechanism. (a) Dual sweep transfer curves of the MoS$_2$ device ($L_{ch}$ = 2 μm) are plotted against temperatures at 250 K, 300 K, and 450 K. The gate voltage sweep range is taken from -60V to 40V and again back to -60V with a sweep rate of 0.5 V/s; $V_{ds}$ is fixed at 100 mV. The forward (backward) sweep is represented by solid red (blue) lines indicating a clockwise hysteresis loop marked by arrows. Cross-sectional schematic representations of the device portray the possible interfacial mechanism that involves (b) charge-(de)trapping processes in clockwise hysteresis at high*

*negative ($V_g<0$) and positive gate voltages ($V_g>0$). (c) the movement of mobile ions in gate oxide under (f) high negative ($V_g<0$) and (g) positive gate voltages ($V_g>0$). (d) Dual sweep hysteresis behaviors on CVD grown MoS$_2$ synthesized using conventional techniques (without NaCl). The transfer curves show the emergence of clockwise hysteresis with respect to temperature. As shown, the degree of hysteresis increases maintaining the same direction (clockwise) as we increase the temperature from 250 to 450 K.*

The transient drain current analysis provides an elementary route to understand subsequent intermediate mechanisms at different time scales that alter with temperature under positive bias at the gate terminal. Figure 3 (a-d) represents the waveform nature of the drain current by employing a single pulse gate voltage at four different temperature regions starting from 300 to 450 K. Each pulse waveform has a period of ~12s which consists of three components of equal intervals; an initial READ ($R_0$) at $V_g = 0$ V, WRITE (W) at $V_g = 30$ V followed by a final READ ($R_1$). A constant $V_{ds}$ of 300 mV is used for all the above operations. The insets of each figure represent the magnified view of the WRITE process and its progression with temperature. At 300 K, READ provides initial drain current values, which increase suddenly as the pulse amplitude increases from 0 to 30V due to the accumulation of charge carriers into the MoS$_2$ channel, as shown in Fig. 3 (a). However, during the WRITE process, while the voltage pulse is maintained at 30 V, the drain current drops quickly, followed by a slight decrease attributed to two types of charge trapping, i.e., fast and slow at the MoS$_2$/SiO$_2$ interface. This degradation of drain current due to positive gate stress is well consistent with previously reported results.[25] The final READ ($R_1$) process provides nearly identical drain current values as that of initial READ ($R_0$) such that the ratio between them, i.e., $R_1/R_0$ remains near unity. The drain current reduction during WRITE operation is best fitted with two trap model by employing the following equation;

$$I = I_0 + A\left(e^{\frac{-(t-t_0)}{\tau_1}}\right) + B\left(e^{\frac{-(t-t_0)}{\tau_2}}\right) \qquad (1)$$

where $\tau_1$, and $\tau_2$ are time constants for fast and slow trapping processes, respectively, A and B are initial current amplitudes, and $I_0$ and $t_0$ are adjustable offsets. At 350 K, the fleeting drain current during the WRITE operation shows a near similar behavior, except a slow increase of current values can be observed after an initial degradation of current, as shown in Fig. 3 (b). In this case, the $R_1/R_0$ ratio still provides near unity value. However, when the temperature exceeds 350 K, i.e., at 400 K, the drain current during the WRITE process reforms its behavior and is observed to increase sharply till the gate stress is applied. Interestingly, as shown in Fig. 3 (c), the $R_1/R_0$ ratio also increases to 1.5, providing an initial signature for possible memory applications of our MoS$_2$ FETs. When the temperature is further increased to 450 K, a more pronounced increment in the drain current values is observed during the WRITE operation, as shown in Fig. 3 (d). In this case, the enhancement in $R_1/R_0$ ratio to 3.57 indicates thermal modulation of drain currents in the presence of gate bias. It is worth mentioning that the sharp increment of current during the WRITE process always follows a very short-lived degradation of drain currents (as indicated by an arrow in the inset), which uncovers two possible mechanisms that are taking place at high temperatures. The origin of the initial degradation of drain currents is previously discussed, which corresponds to the trapping of charge carriers at the MoS$_2$/SiO$_2$ interface. However, the trend is carefully analyzed to shed light on the mechanism of the increased drain current. We notice that the increasing trend of drain current closely resembles a diffusion-like current and is accordingly best fitted with the following current monitoring equation,[34]

$$I = I_0 - A(e^{-\frac{D\pi^2}{L^2}t}) \qquad (2)$$

where, $I_0$, and A are standard derived parameters, L is the channel length, and D is the diffusion coefficient. The temperature-dependent progression of underlying mechanisms is shown in Fig. S5 (see supplementary Information), where the left Y axis represents the trap time constants (outcome of trap model), and the right Y axis represents the diffusion coefficient (outcome of current monitoring model). As the temperature rises, the two trap constants (fast and slow) initially provide finite values and then become negligible around 400 K. However, the diffusion coefficient has a finite value above 400 K, stipulating its dominant nature at high temperatures. The charge trapping process is more favourable at room temperature, quenched by the migration of ions near the $MoS_2/SiO_2$ interface at high temperature. Here, it is worth mentioning that the accumulation of ions enroute the memory capabilities in ML $MoS_2$ FETs, which will be further discussed in the following sections.

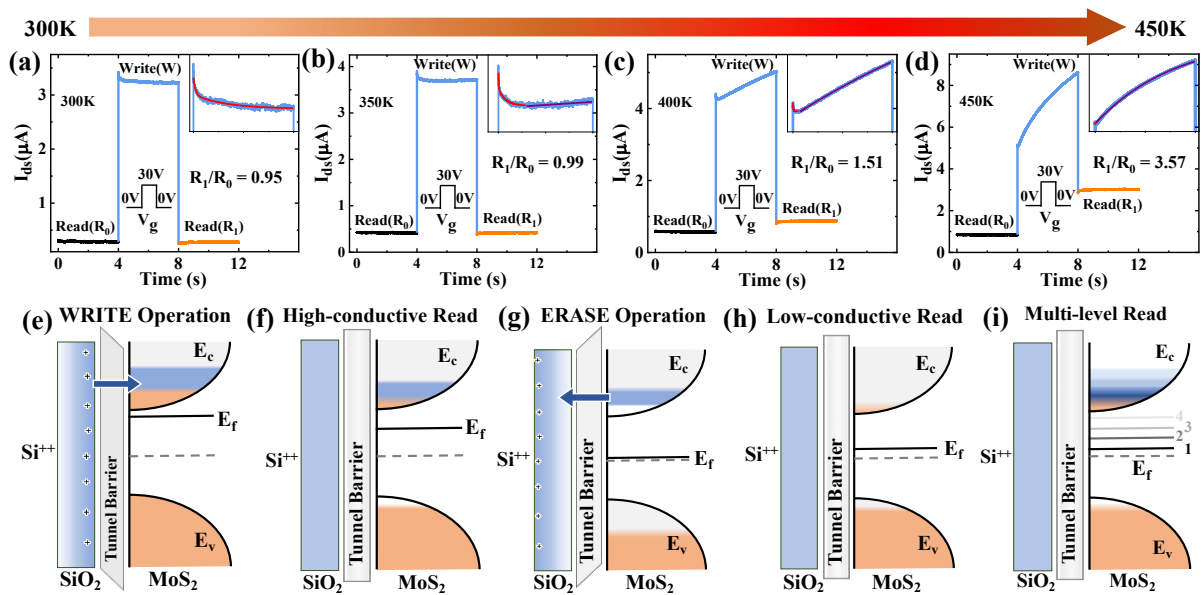

*Figure 3. Transient drain current analysis and proposed energy band diagrams (a-d) represent the waveform nature of the drain current by employing a single pulse gate voltage at four different temperature regions starting from 300K to 450K. Each pulse waveform has a period of ~12s which consists of three components of equal intervals; one is at $V_g$ = 30 V, known as WRITE (W) operation, and two are at $V_g$ = 0V, known as READ ($R_0$ and $R_1$) operations. The drain-source voltage is kept fixed at 300 mV. The insets of each figure represent the magnified view of the WRITE process and its progression with temperature. The*

*fittings to the fleeting drain currents during the WRITE operations are represented by red and magenta colors for the two trap models and the current monitoring model, respectively. The $R_1/R_0$ ratio is calculated for each temperature case. The proposed mechanism with suggested band diagrams of MoS₂/SiO₂/Si to explain the single and multi-level memory effects in MoS₂ FETs. (e) WRITE operation at $V_g > 0$. (f) High conductive READ state at $V_g = 0$. (g) RESET operation at $V_g < 0$ (h) Low conductive READ state at $V_g = 0$. The orange filled color represents the carrier accumulation during the transistor switch-on, and the blue filled color represents the carrier tunneling due to the accumulation of active ions at the MoS₂/SiO₂ interface. Arrow represents the tunneling direction. (i) Multi-level READ state at $V_g = 0V$. Different color bars represent multiple charge carrier injections at every implemented programmed pulse cycle and the successive elevation of Fermi levels.*

To explain the non-volatile memory states in MoS₂ FETs originated by the thermally-driven accumulation of ions at the MoS₂/SiO₂ interface, we propose a hypothesized mechanism involving WRITE, READ, and ERASE schemes. Figure 3 (e-h) demonstrates the energy band diagrams for different gate voltage polarities to acknowledge basic n-type transistor functions and extrapolation of programmable conductance states for both single and multi-level memory capabilities. As shown in Fig. 3 (e), when a high positive voltage ($V_g > 0$) is applied to the gate terminal of the MoS₂ FETs, the Fermi-level shifts up, and accumulation of carriers (shown in orange) in the conduction band of MoS₂ essentially switch on the transistor offering a n-type FET behavior. Additionally, as the device is characterized at elevated temperature, the thermally induced ions drift and migrate towards MoS₂/SiO₂ interface due to high positive gate bias. Many accumulated ions persuade high carrier densities at the interface during positive gate stress. As a result, lowered tunneling barrier favors injecting additional charge carriers (shown in blue) into the conduction band of MoS₂. It should be noted that this injection process continues till the positive gate voltage is maintained, which constantly populates the conduction electrons in the MoS$_2$ channel. This behavior is consistent with the observed results in Fig. 3 (d), where the drain current progressively increases during the pulsed gate voltage maintained at 30V. This is denoted by the WRITE process, where conductance states can be achieved by allowing continuous migration and accumulation of ions near the SiO$_2$/MoS$_2$ interface, increasing the carrier densities in the MoS$_2$ channel over a progressive period by applying a positive gate voltage pulse. Figure 3 (f) demonstrates the band diagram for the high conductivity read state after immediate withdrawal of the gate field (V$_g$ = 0V). The Fermi level is lowered as the transistor goes to the off state at zero gate bias. However, the injected carriers mediated by the ions have a low probability of tunneling back through the barrier. As a result, despite the transistor being in the off state at zero gate bias, a finite density of carriers (shown in blue) is retained in the conduction channel of the FET, giving rise to a high conductivity state after the writing process. When a high negative bias is applied to the gate terminal (V$_g$ < 0), the transistor remains in the off state, and the Fermi level is further lowered. However, as shown in Fig. 3 (g), the accumulated ions diffuse and migrate away from the MoS$_2$/SiO$_2$ interface. This is known as ERASE process, where the additional injected carriers are removed from the conduction channel at a high negative gate field, regressing the device to the original neutral state. The read process is further performed at V$_g$ = 0V, giving rise to a low conductivity state, where the carriers mediated by ions don't contribute to the resultant drain current, as shown in Fig. 3 (h). Apart from single memory operation, which consists of a pair of (high and low) conductive read states, the proposed mechanism could adequately explain the multi-level memory effects arising in our MoS$_2$ FETs. As illustrated in Fig. 3 (i), successive WRITE operations lower the barrier allowing the charge carriers to tunnel into the conduction band of MoS$_2$, independently increasing the carrier densities and lifting the Fermi level at each pulse cycle. Continuing this process generates higher conductance states where each state corresponds to pulse cycle biasing history. This way, we could observe multi-level memory effects in our MoS$_2$ FETs.

The development of a novel thermally-driven and ion-mediated non-volatile multi-level storage capacity can further be realized owing to the anti-clockwise hysteresis nature of our MoS$_2$ FETs at high temperatures. Figure S6 (see supplementary Information) shows dual sweep transfer characteristics with varying hysteresis widths for 10 successive cycles at four different temperatures. In each cycle, the gate voltage is swept from -10 to 40 V and back to -10 V. With repeating cycles, the hysteresis curves initially provide overlapping current loops at 300K. However, the current loops shift upwards at elevated temperatures with each successive cycle, particularly at 475K. This illustrates multiple conductance states can be achieved corresponding to several dual sweep cycles, which spurred the attainment of multi-level memory generation of our MoS$_2$ FETs. Figure 4 (a) shows multi-level programmed state operations driven by gate pulse trains having pulse width 1s. Based on our single-state memory function at 450 K, a positive gate pulse train (10 cycles) with an amplitude of 30V is provided for multiple writing processes, as shown in the lower panel (blue waveforms) of Fig. 4 (a). The reading of distinct conductance states is performed at $V_g = 0$ V after each writing process establishing a multi-level storage unit in the case of ML MoS$_2$ FET. Note that a complete RESET of the device is necessary (with $V_g = -70$ V) before the initial writing process to recover from any higher conductance state (blue bar). The drain-source voltage is kept fixed at 300 mV at every step of the programmable memory operations. The upper panel (red waveforms) shows the fleeting drain currents during RESET, WRITE, and READ programs. The green gradient bars (a guide to eyes) represent the individual READ processes after periodic WRITE operations, which assimilate increased distinct conductance states with several pulse cycles. The READ current rises progressively with increasing the pulse number, a phenomenon that represents the continual accumulation of carriers in $MoS_2$ as prolonging the gate stress on the memory.

To further explore the pulse-controlled memory behaviour, the multi-level memory cycles are expanded to different programming pulsed width measurements, attributed to distinct conductance states based on the number of cycles. Figure S7 represents such memory cycles for pulse width 2, 3, and 4s, respectively. It is worth noticing that using the programmed memory operations with longer pulse widths yields higher conductance states, which reflects that n-bit memory can be achieved by varying the pulse time. Additionally, we witness an interesting result on data erasing, i.e., the multi-level storage information can be erased gradually by applying low negative voltages accompanying the complete RESET at a high negative bias (-70 V). Figure 4 (b) represents a multi-level ERASE operation with programmable erase pulse train of pulse width 4s where the ERASE and READ processes are performed at -40 and 0 V, respectively (see the voltage waveform in the lower panel). Similar to the multi-level WRITE operation, a progressive reduction of the READ current with successive ERASE operations indicates continual decumulation of charge carriers in the conduction channel. This means any desired lower conductance state can be achieved from a higher conductance state by performing ERASE operations with n-cycle pulse trains. The multi-level WRITE and ERASE operations in our device show exceptional pulse control of charge injection and release, providing a complete yet simplest way to achieve large-scale data storage capabilities in $MoS_2$ FETs. The increment and decrement of current updates with the number of pulses provide a potential platform to achieve high temperature in-memory neuromorphic computing for online learning, discussed subsequently.

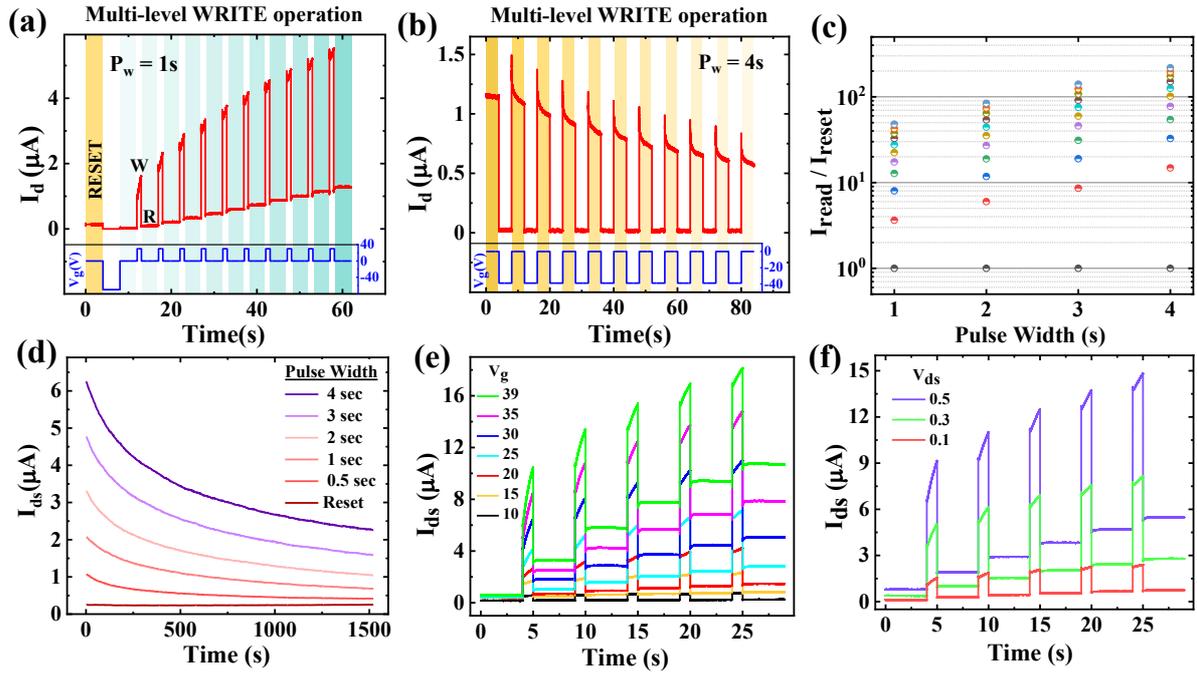

*Figure 4. Dynamic responses of multi-level ionotronic memory* *(a) Multi-level programmed state operations at 450K driven by gate pulse trains with pulse width 1s. A Positive gate pulse train (10 cycles) with an amplitude of 30V is provided for multiple writing processes, as shown in the lower panel (blue waveforms). The upper panel (red waveforms) shows the fleeting drain currents during RESET, WRITE, and READ programs. The green gradient bars represent the individual READ processes after periodic WRITE operations, which assimilate to increased distinct conductance states with several pulse cycles. (b) Multi-level ERASE operation with a programmable pulse train of pulse width 4s where the ERASE and READ processes are performed at -40V and 0V, respectively (see the voltage waveform in the lower panel). (c) READ-to-RESET ratios for 10 successive pulses of widths 1s, 2s, 3s, and 4s. (d) The retention curves of the observed memory behavior after 3 successive WRITE operations for 1500 s with various pulse width conditions. Multi-level programmed state operations at 450 K driven by gate pulse trains (e) with gate pulse amplitude varied from 10 to 40V (f) with drain-source voltage for 0.1, 0.2, and 0.3V.*

The thermally-driven multi-level memory functions discussed above can generate large set of data storage possibilities by controlling the pulse trains of different pulse widths with n-successive cycles. The pulse-dependent distinct READ states are shown in Fig. S8 (see supplementary information). The READ-to-RESET ratio, a desirable feature of a memory device, is calculated for all the pulse-controlled READ states shown in Figure 4 (c). For each pulse width, the ratio values are plotted in logarithmic scale for 10 successive pulses represented by colored half-filled circles spanned over a large range from 1 to 220. The ratio value 1 is represented by the OFF-state in all pulse widths. The ON/OFF ratio for the first pulse varies from 3 to 16 as the pulse width varies from 1s to 4s. Interestingly, the pulse width with longer time has higher and distinct READ window values showing strong dependence on various pulse conditions for each corresponding cycle. An attempt has been made to realize the ON/OFF ratios in thermally-aided mem-transistors in 2D materials; for example, Goyal *et. al.* reported READ windows of about 1.9 and 7.4 in few-layered $MoS_2$ and $ReS_2$ memory devices, respectively, for single-step memory operations.[35] Our results show unambiguously higher READ values in the case of ML $MoS_2$ FETs. More importantly, our studies show multi-level memory states in our device can take this ratio up to ~220 and beyond by using longer pulse trains with varying pulse widths. It is also worth noting that combinational pulse waveforms can get desired READ window and store information in multiple storage cells. The retention curves of the observed memory behavior are obtained after 3 successive WRITE operations for various pulse width conditions, shown in Fig. 4 (d). The retention behavior shows that the data levels of different pulse conditions are well-discernable after 1500s. It may also be noted that our device shows much better retention capabilities with increasing pulse widths. Such memory devices can be used in electronic components that require low-scale information storage functionalities such as cache memory, disposable electronic tags, buffer memories, etc.

Apart from various pulse conditions, i.e., several pulses and pulse width, multi-level data storage can be obtained by changing the gate voltage amplitudes. In previous reports using floating gate memory configurations, optical response, and plasma treatment, the gate voltage-dependent multi-bit generation is achieved in 2D materials and heterostructures.[19,36–39] Here, we also find a strong dependence on the drain current levels as we increase the $V_g$ from 10V to 40V, as shown in Fig. 4 (e). The programmable memory operations are obtained for 5 successive pulses, which represent a similar progressive increase in READ states, i.e., the higher gate voltage amplitude corresponds to larger READ windows. This indicates gate voltage amplitude is crucial for obtaining stable and distinct memory states. We are also keen to observe the effects of drain-source voltage, i.e., $V_{ds}$, on the memory operations in the $MoS_2$ FETs, shown in Fig. 4 (f). All these above experiments strongly vow the versatile nature of our memory device, which can be tuned with different controllable parameters such as the number of pulses, pulse-width, pulse-amplitude, and drain-source voltage.

As discussed earlier, non-volatile memory (NVM) devices present a potential application as an in-memory computing element,[40] which can be accomplished via the multi-level conductance response of an NVM, providing the capability to store analog synaptic weights of an Artificial Neural Network (ANN) on-chip.[41] ANNs draw inspiration from the neural connectivity of the human brain, although they do not align with any particular biological learning paradigm. Notably, in-memory computing is capable of executing matrix-vector multiplication (MVM) operations, which are the primary computations utilized in the domain of Artificial Intelligence (AI).[42] According to this approach, a crossbar array of NVM devices can undertake the MVM operation by encoding the input vector as an analog voltage and the weight matrix as analog conductance values stored in the memory devices. It may be worth mentioning that the main focus of most of the neuromorphic computing literature is based on conductance weight updates at room temperature for online learning.[20,22,43] As the field of neuromorphic computing is growing rapidly, it is crucial to shed light on its high temperature applications. However, a few attempts have been made to understand high temperature neuromorphic learning aspects in integrated synaptic devices.[44,45] In this regard, by tuning the presynaptic voltage pulses, we demonstrate near linear synaptic weight updates for high temperature neuro-inspired online learning behavior in our $MoS_2$ based memtransistor, and the results are discussed in the following section.

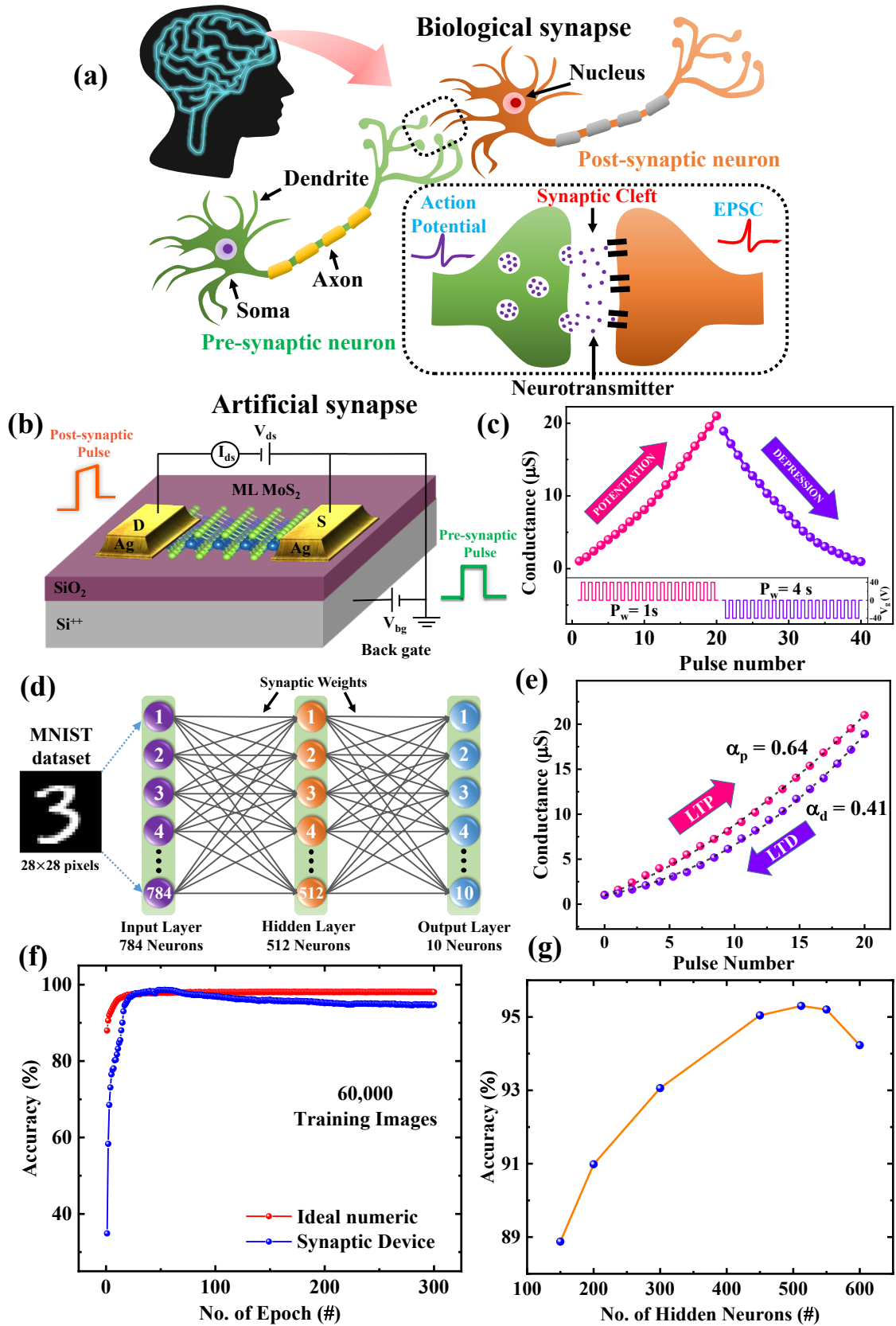

*Figure. 5 Artificial synaptic learning of MoS$_2$ based ionotronic devices* *(a) Demonstration of biological synapse operation between pre and post-synaptic neurons generating EPSC. (b)*

*Schematic illustration of our proposed MoS₂ based artificial synapse in accordance with biological response. (c) Long-term plasticity characteristics of the proposed device emulating long term potentiation (LTP) and depression (LTD) with 20 consecutive identical pulses of* (40V, 1s), *and* (-40V, 4s), *respectively. The* $V_{ds}$ *of 0.1V is applied to read the change in EPSC. (d) The schematic diagram of a three-layer artificial neural network with 784 input nodes, 512 hidden nodes, and 10 output nodes for the classification of MNIST 28×28 pixels. (e) Extraction of nonlinearity factors for LTP and LTD, (f) Training accuracy evolution (%) vs. the number of epochs for the software-based ideal numeric case and the proposed ion-driven synaptic device during NN training. (g) Training accuracy evolution (%) vs. the number of hidden nodes ranging from 100 to 600.*

A biological synapse in the human cerebrum transmits neurological impulses through electrical and chemical signals between pre and post-synaptic neurons. Neurotransmitters released from the axon of the presynaptic neuron attach to neuro-receptors at the dendrite of the postsynaptic neuron, generating an excitatory postsynaptic current (EPSC), as illustrated in Fig. 5 (a). The magnitude of EPSC depends on the synaptic weight, which can be modified by the presynaptic spike. This is critical for memory and cognitive processing.[21] The proposed ionotronic device based on ML MoS₂ can function as an artificial synaptic transistor, emulating the biological synapse, as presented in Figure 5 (b). MoS₂ plays the role of the synapse, where the back-gate voltage is considered as the presynaptic stimulation, and the drain/source electrode acts as the postsynaptic terminal responsible for accumulating EPSC. Figure 5 (c) depicts the change in channel conductance with the number of repetitive presynaptic stimulation pulses. To imitate long-term synaptic plasticity in our device, identical potentiation (consisting of voltage pulses with amplitude of 40V and width of 1s) and depression (consisting of voltage pulses with amplitude of -40V and width of 4s) pulses are consecutively implemented to the gate terminal for 20 times as a presynaptic spike. To perceive the change in channel conductance, a bias ($V_{ds}$) of 0.1V is applied. As shown in Figure 5(c), our proposed synaptic transistor shows remarkable linear and symmetrical conductance updates. Notably, the dynamics of potentiation and depression, specifically the dynamic range, linearity, and asymmetry, play a pivotal role in ensuring the precision of learning and recognition simulation.[46] The outstanding linearity in our case could provide a potential opportunity to achieve high classification accuracy in ANN training. We have attained a dynamic range of conductance ratio [$G_{Max}/G_{Min}$] equal to ~20, which may be deemed as the on/off ratio value for accomplishing superior performance in ANN tasks. We confidently ascribe the enhanced linearity of the conductance updates to the ion-driven charge transfer, storage, and release. Additionally, we employ a synaptic model to replicate the LTP and LTD features demonstrated in Fig. 5(e) below[47]

$$G = \begin{cases} ((G_{Max}^\alpha - G_{Min}^\alpha) \times \omega + G_{Min}^\alpha)^{1/\alpha} & if \ \alpha \neq 0 \\ G_{Min}^\alpha \times (G_{Max}/G_{Min})^\omega & if \ \alpha = 0 \end{cases} \quad (3)$$

The variable ω and the maximum ($G_{Max}$) and minimum ($G_{Min}$) conductance determine the non-linearity coefficients ($\alpha$) for potentiation ($\alpha_p$) and depression ($\alpha_d$), which are 0.64 and 0.41, respectively. These parameters play a crucial role in the off-chip training procedure for pattern recognition in neural network training.[48]

To verify the image classification performance of our synaptic device, we simulate an ANN using the experimentally determined long-term plasticity characteristics to facilitate supervised learning of large image handwritten digits (28×28 pixels) from the Modified National Institute of Standards and Technology (MNIST) dataset.[49] An open-source software (Pytorch) is utilized for implementing the simulations.[50] The training images are divided into batches of 32, with each batch image linearized to a 784×1 input matrix of pixel intensities normalized to [0, 1]. A total of 784 input nodes were then linked to 512 hidden nodes, which, in turn, are linked to 10 output nodes, as depicted in Fig. 5 (d). These 10 output nodes correspond to the output classes of the MNIST dataset, which are digits ranging from "0" to "9". The NN training algorithm used the rectified linear unit activation function and cross entropy loss as the cost function with a back propagation method. The training session is repeated five times for 300 epochs, and the mean values for the training accuracy are plotted as a function of the training epochs, as shown in Fig. 5 (f). Our synaptic device achieves a maximum training accuracy of ~98% and eventually drops to a stable limit of ~95% for the pattern recognition task at the end of the training cycles, which strongly suggests the neuromorphic adaptation of our artificial synaptic transistors. Figure 5 (g) displays the relationship between the number of hidden nodes and accuracy at 200 epochs. As shown in the graph, an increase in hidden nodes leads to a higher recognition rate, with a maximum rate achieved at 512 nodes. These findings indicate the potential of our ion-driven $MoS_2$-based transistors for building multilevel in-memory synapse arrays, facilitating complex data processing tasks.

## 3. Conclusions

In summary, we have demonstrated a novel functional three terminal device with salt-assisted CVD-grown ML $MoS_2$ as channel material which serves as a high geared intrinsic transistor at room temperature and becomes a multi-level memory cell at relatively high temperature. The memory behavior is attributed to the hysteresis collapse and switch from clockwise to anticlockwise, which further expands with increasing temperature. The reverse hysteresis is ascribed to the migration of active ions and accumulation at the $MoS_2/SiO_2$ interface due to electrostatic gating at a higher temperature of around 450 K. The proposed thermally-driven programmable memory operations, i.e., READ, WRITE, ERASE, are found to be modulated by gate voltage biasing history. The step-like READ-RESET ratio and retention curves are obtained for various pulse conditions, which stipulates vigorous data storage capabilities in thermally-driven memory cells. Furthermore, the multi-level memory states are investigated with varying pulsed width, gate voltage amplitudes, and drain-source voltages. We corroborate these results with the relative energy band diagrams to explain the single and multi-level memory effects in MoS$_2$ FETs. Moreover, our device demonstrates excellent linearity and symmetry upon electrical potentiation and depression for high-temperature synaptic applications. As a result, a high classification accuracy of 95% is achieved during training and testing of the Modified National Institute of Standards and Technology (MNIST) datasets in artificial neural network (ANN) simulation. Our work serves as a precursor to novel pathways in the realm of 2D semiconductors, aimed towards achieving robust high temperature in-memory computing applications with enhanced memory capabilities for artificial cognitive development after the human brain.

**Methods**

**Synthesis of monolayer MoS$_2$:** The monolayer MoS$_2$ is grown using a salt-assisted CVD synthesis as reported elsewhere.[25] However, some growth parameters are varied from the previous literature, not only to achieve large scale triangular domain growth with high crystallinity but also to diffuse mobile ions into the SiO$_2$ gate dielectric. More specifically, the amount of the salt precursor, i.e., NaCl, is increased to promote the nucleation density, enhancing the lateral dimension of MoS$_2$ domains. Additionally, the precursor (MoO$_3$+NaCl) height is increased so that the distance between the precursor and growth substrate is minimized to 3 mm to channelize the sulfur feeding during growth time.

**Device Fabrication:** The monolayer MoS$_2$ based mem-transistors are fabricated by using a photolithography (Heidelberg μPG 101) system. Initially, the salt-assisted CVD grown MoS$_2$ samples on SiO$_2$/Si substrate are coated with a positive photoresist (ma-p-1205) by a spin coater (SUSS Microtech) and then baked at 80 °C for 1 min. Under the inspection of a high-resolution microscope, the contact patterns of several channel lengths are exposed on the monolayer MoS$_2$ flakes with a 405 nm laser in photolithography. The exposed patterns are developed with an alkaline solution (1:4, NaOH:DI water) for 1 min. The sample is then mounted on a thermal evaporation chamber for the deposition of the silver (Ag) electrode, followed by the dissolution of residual resists in acetone for 10 mins, known as lift-off process. The heavily p-doped silicon functioned as the gate electrode, and 285 nm $SiO_2$ functioned as the gate dielectric. Before taking all the electrical measurements for this work, the devices are annealed at 200 °C for 20 hours in a high vacuum (~$10^{-6}$ mbar) condition.

**Electrical Characterization:** The fabricated device is mounted on a cryogenic four-probe station (Lake Shore) to probe the top and bottom electrodes. The chamber is maintained at a high vacuum of ~$10^{-6}$ mbar during all the measurements. For electrical characterization, a semiconductor parameter analyzer system (Keithley 4200A-SCS) is used. A high-speed pulse generator module 4220-PGU and measurement unit 4225-PMU integrated within Keithley 4200 system are used for electrical pulse characterization. In order to prevent photo-excitation of charge carriers, all experiments are performed in dark conditions.

**Raman spectroscopy and PL mapping:** The Raman spectra are collected with a confocal micro-Raman spectrometer (Renishaw Invia) using a laser excitation wavelength of 532 nm in a backscattering configuration employing a 100 Å~ (NA = 0.8) objective. The laser power on the sample was kept low to avoid local heating. The laser exposure time on the sample is kept fixed for 10s with 2 accumulations. However, for PL mapping, the exposure time is reduced to 1s to avoid local heating during the spectral acquisition.

## Supporting Information

Supporting Information is available free of charge which contains additional results of device characterizations.

## Acknowledgements

S.P.S. thanks to the Science and Engineering Research Board for financial support (SERB/F/7481/2020-2021). The authors acknowledge the use of the Micro-Raman facility at the central research facility (CRF) of KIIT Deemed to be University, Bhubaneswar.

## Conflict of Interest

The authors declare no competing financial interest.

## Data Availability Statement

The data that support the findings of this study are available in the supplementary material of this article and from the corresponding author upon reasonable request.

## Keywords

High temperature transport; Reverse hysteresis; Monolayer $MoS_2$ transistors; Multi-level Non-volatile memory; Neuromorphic Computing